**Graphene in 2D/3D Heterostructure Diodes for High Performance Electronics and Optoelectronics**

*Zhenxing Wang, Andreas Hemmetter, Burkay Uzlu, Mohamed Saeed, Ahmed Hamed, Satender Kataria, Renato Negra, Daniel Neumaier, and Max C. Lemme\**

Dr. Zhenxing Wang, Andreas Hemmetter, Burkay Uzlu, Prof. Daniel Neumaier, Prof. Max C. Lemme
AMO GmbH, Otto-Blumenthal-Str. 25, 52074 Aachen, Germany
E-mail: lemme@amo.de

Andreas Hemmetter, Burkay Uzlu, Dr. Satender Kataria, Prof. Max C. Lemme
RWTH Aachen University, Chair of Electronic Devices, Otto-Blumenthal-Str. 2, 52074 Aachen, Germany

Dr. Mohamed Saeed, Dr. Ahmed Hamed, Prof. Renato Negra
RWTH Aachen University, Chair of High Frequency Electronics, Kopernikusstr. 16, 52074 Aachen, Germany

Prof. Daniel Neumaier
University of Wuppertal, Chair of Smart Sensor Systems, Lise-Meitner-Str. 13, 42119 Wuppertal, Germany

Keywords: graphene, metal-insulator-graphene diodes, integrated circuits, high frequency, Schottky diodes, photodetectors, infrared

Abstract:

Diodes made of heterostructures of the two-dimensional (2D) material graphene and conventional three-dimensional (3D) materials are reviewed in this manuscript. Several applications in high frequency electronics and optoelectronics are highlighted. In particular, advantages of metal-insulator-graphene (MIG) diodes over conventional metal-insulator-metal diodes are discussed with respect to relevant figures-of-merit. The MIG concept is extended to one-dimensional diodes. Several experimentally implemented radio frequency circuit applications with MIG diodes as active elements are presented. Furthermore, graphene-silicon Schottky diodes as well as MIG diodes are reviewed in terms of their potential for photodetection. Here, graphene-based diodes have the potential to outperform conventional photodetectors in several key figures-of-merit, such as overall responsivity or dark current levels. Obviously, advantages in some areas may come at the cost of disadvantages in others, so that 2D/3D diodes need to be tailored in application-specific ways.

# 1. Introduction

Diodes are basic electronic devices and essential building blocks of modern (opto-)electronics. They find applications in different areas, for example as radiofrequency demodulation, power rectification or over-voltage and reverse-voltage protection, and others.[1–3] Diodes can furthermore function as light detectors, i.e. photodiodes. In conventional semiconductor technology, the most widely used diodes are p-n junction diodes and Schottky diodes.[1–4] Thin-film metal-insulator-metal (MIM) diodes are an alternative concept for specific applications. The fabrication of MIM diodes is rather simple and established, which enables applications where classic semiconductor diodes cannot be integrated directly or where they do not provide the required performance.[5–8]

Graphene has been intensely researched in the electronic device community for over a decade due to its excellent electronic and optoelectronic properties, including applications as transistors and diodes [9–11]. Although graphene enables transport of both p- and n-type charge carriers, the gap-less band structure as well as Klein-Tunneling lead to a linear I-V characteristic of lateral p-n junctions in graphene[12,13]. As a consequence, graphene field effect transistors are not suitable for logic circuits with high on-off ratios [9]. In optoelectronics, in contrast, lateral graphene p-n junctions can be utilized to enhance photogenerated currents with the photothermoelectric effect [14–16]. However, two other kinds of graphene-based diodes have been realized in the past years: metal-insulator-graphene (MIG) diodes, in which current is governed by either tunneling or thermionic emission across an insulating barrier, and graphene-silicon (Si) Schottky diodes, in which a Schottky barrier is formed at the graphene-Si interface. In this review, we focus on these two types of graphene-based diodes and compare their performance with devices made in conventional technology. Furthermore, we present their application in RF electronics and circuits, as well as in optoelectronics. Beyond MIG and graphene-Si diodes, there are many other types of graphene based 2D/3D heterostructures such as graphene-germanium or graphene-gallium nitride photodetectors,[17–19] or van der Waals heterostructures based entirely on 2D materials operating as photodetectors, electronic devices or solar cells.[20–24] However, these other material systems are beyond the scope of this review article.

## 2. Operating Principles of Metal-Insulator-Graphene and Graphene-Silicon Schottky Diodes

Here, we introduce the band structures of the MIG junction and the graphene-silicon junction to explain the basic operation principles of these diodes.

### 2.1. Metal-Insulator-Graphene Junctions

A fundamental limiting factor in the development of diodes with high frequency response is their charge carrier transport mechanism.[25] Currents in semiconductor p-n junctions are drift currents that form due to an applied electric field. Minority charge carriers take a finite time to cross the junction and charges accumulate in the depletion region. When the bias is reversed rapidly, the depletion region is first discharged and thus briefly conductive. This parasitic diffusion capacitance limits the reverse recovery time of p-n junction diodes to over 10 ns, limiting the high frequency application range.

MIM diodes overcome this limitation by reducing the charge that is accumulated between the terminals. An MIM diode consists of a stack of two metal contacts separated by a thin (several nanometers) insulator.[26] The metals are chosen to have different work functions to each other and the insulator needs to have an electron affinity that is lower than the work function of either contact metal. The blue trapezoids in Figures 1a to 1c show the barrier shape of an MIM junction schematically at different biases. In thermal equilibrium, the band structure of this MIM stack forms an asymmetric tunnel barrier. Electrons can be transported through the potential barrier (tunneling) or over the barrier (thermionic emission). Since the work function of both metals remains constant, only the applied bias controls the shape of the barrier, which in turn determines the dominant transport mechanism.[27] Important figures-of-merit (FOM) for diodes include current asymmetry ($f_{ASYM} = |J_F/J_R|$), nonlinearity ($f_{NL} = (dJ/dV)/(J/V)$) and responsivity ($f_{RES} = (d^2J/dV^2)/(dJ/dV)$). They can be optimized for specific applications, in competition with the junction resistance of the MIM diode.[28,29] The FOMs are typically improved when the barrier is thick (> 10 nm) and thermionic transport dominates, while the diode on-current increases when tunneling dominates in the case of a thinner barrier (< 5 nm).[27] Replacing one metal with graphene can overcome this tradeoff and simultaneously improve the static performance of such a MIG diode, while maintaining a high overall current level.[27] This is due to the electrostatic tunability of the Fermi level in graphene, which allows tuning not only the barrier shape, as in MIM diodes, but also the barrier height at the graphene-insulator junction (Figure 1d to 1f). In each case the graphene side is grounded. When a positive bias is applied to the metal in the two-terminal MIG diode, the barrier height at the graphene-insulator interface is lowered. Since the thermionic current is determined by the barrier height, the total current in the forward direction is larger than in the case of a MIM diode, where the metal-insulator barrier height remains constant (Figure 1c). A negative bias applied to the metal increases the graphene-insulator barrier height, thus lowering the thermionic current in the reverse direction. This bias-dependent barrier height modulation has been simulated previously[30]. This mechanism enhances the asymmetry of both thermionic and tunneling currents, increasing also the nonlinearity, responsivity and on-currents.[27]

## 2.2. Graphene-Silicon Schottky Diodes

When metal contacts a semiconductor, the difference in the work functions of the two materials leads to electron transfer across the interface. This results in the formation of a built-in potential ($\phi_{bi}$) across the interface due to the band bending in the semiconductor. The junction between a metal and a lightly doped semiconductor is called Schottky junction, and acts like a barrier for charge carriers depending on the type of semiconductor (i.e. *n* or *p*) and biasing conditions. Such a junction is rectifying in nature and allows current to flow only in one direction. Under forward bias conditions, high current flows through the diode ("on state") while in reverse bias conditions, there is negligible current in the device ("off state").

The difference between the work function of the metal and the electron affinity of the semiconductor is known as Schottky barrier height (SBH) and is an important parameter affecting the diode characteristics.

Unlike p-n junction diodes, the current transport in Schottky barrier diodes is controlled by the emission of majority charge carriers. Schottky barrier diodes are used as highly efficient photodetectors due to their higher speed and efficiency compared to p-n junction diodes. In the case of metal-semiconductor Schottky diodes, very thin semitransparent metal films are generally used to allow the penetration of light into the active region. This results in a trade-off between transparency and conductivity of the metal film. Graphene fulfills both criteria in an excellent way, as it is highly conductive and nearly transparent in a wide spectral range from ultraviolet to the infrared.[31] The physics of graphene-Si junctions is similar to metal-semiconductor junctions with the main difference that graphene has a low density of states, and its Fermi level can be modulated electrically or chemically [32–35]. Figure 2 shows the schematic of a graphene sheet contacting an n-type semiconductor at zero bias. Graphene is typically p-doped under ambient conditions and, hence, it forms a Schottky junction with n-type Si or with other semiconductors, and an ohmic contact with p-type Si [36]. In graphene/Si Schottky diodes, the atomically thin and broadband transparent graphene forms a Schottky junction with Si, which allows the efficient exposure of the entire active diode area to light.

## 3. Electronic Devices and Circuits

This section describes the operation of MIG diodes, the extension of the device concept to one-dimensional (1D)-MIG diodes and opportunities for exploiting MIG diodes in integrated circuits.

### 3.1. MIG Diodes

The high carrier mobility in graphene provides a low series resistance in MIG diodes. Combined with a low junction capacitance and an insulator with a low energy barrier, MIG diodes become promising components for radio frequency (RF) applications, potentially into the submillimeter-wave or terahertz (THz) region. A typical MIG diode structure is shown schematically in **Figure 3**a. Early experimental demonstrations have been realized using the low barrier insulator titanium oxide ($TiO_x$) and showed excellent rectification (**Table** 1).[27,30] This diode concept has been used to demonstrate the application as RF power detectors.[30] Here, the metal electrode is buried in the substrate with the stack of the $TiO_2$ insulator and graphene on top. This allows the use of plasma-assisted atomic layer deposition (ALD) to directly deposit a high-quality insulating barrier on the metal, which reduces the parasitic capacitance between the contacts. Temperature-dependent measurements of the current density in Figure 3b show a clear temperature dependency, indicating that thermionic transport dominates. These *I-V* curves can be used to extract the DC diode FOMs asymmetry, nonlinearity, and responsivity. As discussed, MIM diodes commonly show a tradeoff between on-current and static DC performance. The MIG diodes shown here reach a maximum current density of 7.5 $Acm^{-2}$, compared to 2.0 $Acm^{-2}$ for a $Nb/Nb_2O_5$(5nm)/Pt MIM diode, which is the state-of-the-art performance with respect to current density for MIM diodes [30,37] At the same time, the MIG diodes also outperform $Nb/Nb_2O_5$(5nm)/Pt MIM diodes in terms of responsivity with a typical value of 26 $V^{-1}$ for MIG diodes compared to 16.9 $V^{-1}$ of MIM devices, asymmetry (520 compared to 9.8) and nonlinearity (15 compared to 8.2). [30,37] If the thickness of the $Nb_2O_5$ increases to 15 nm, the current level decreases drastically, although other FOMs are enhanced.[7] A detailed comparison of all relevant diode characteristics between MIG diodes and state-of-the-art MIM diodes is shown in Table 1. The static performance data thus demonstrate that MIG diodes can circumvent the tradeoff faced by MIM diodes. In addition, the RF power detection characteristics were measured with a network analyzer as signal source up to 50 GHz. The responsivity reaches 2.8 VW$^{-1}$ at 2.4 GHz and 1.1 VW$^{-1}$ at 49.4 GHz (Figure 4a and 4b).[30] Finally, MIG diodes have also been demonstrated on flexible substrates [38], opening up the possibility for wearable graphene-based RF electronics.

### 3.2. 1D MIG Diodes

Conventionally stacked MIM and MIG diodes do not suffer from a diffusion junction capacitance, yet they nevertheless show a large intrinsic junction capacitance due to their parallel-plate geometry. Furthermore, electrons in MIG diodes are emitted out of the graphene plane and must overcome the "van-der-Waals gap" between graphene and the insulator, which are not covalently bonded and may be further affected by intercalated molecules at the interface. Both effects reduce the maximum operating frequency of such diodes. Scaling of the device area does not improve the cutoff frequency, since junction capacitance and access resistance have opposite proportionality with respect to area.

Carrier emission from graphene edges is known to be very effective in reducing electrical contacts between graphene and metals.[39–41] In a similar fashion, electron emission from the graphene into / across the insulator in MIG diodes can be increased considerably if it happens only through a one-dimensional graphene edge. In this geometry, shown in Figure 3d, the junction area is determined only by the monoatomic thickness of graphene of approximately 0.3 nm and the width of the contact, resulting in an exceptionally small junction capacitance.[42] Simultaneously, the device is able to fully exploit the high charge carrier mobility of graphene, since the current injection occurs in the graphene plane, rather than perpendicular to it. Furthermore, the thin, on-dimensional graphene sheet enhances the electric field at the junction, potentially further improving the charge carrier injection into the insulator. These combined effects are expected to increase the cutoff frequency of 1D MIG diodes.

The DC characteristics of the 1D MIG diodes are comparable to their 2D MIG counterparts (Figure 3c and 3f). Their on-currents, however, are roughly two orders of magnitude higher (Figure 3e), which can be attributed to their lower junction capacitance and series resistance. To compare different devices, it is common to normalize the current to the effective area of the cross section. Since the current is emitted only from the 1D graphene edge, the resulting current densities of the 1D MIG diodes are approximately six orders of magnitude higher than in the case of the 2D MIG diodes.

1D MIG diodes outperform vertical MIG diodes in terms of RF cutoff frequency and current level, with a bandwidth up to 100 GHz.[42] A theoretical cutoff frequency of 2.4 THz has been deduced from the numerically determined junction capacitance and experimental values for the access resistance for a 200-μm long channel, which is achievable by standard photolithography processes.[42] The entire fabrication process is back-end-of-line compatible, scalable and can be carried out in thin-film technology, including on flexible substrates. This opens opportunities for the realization of wearable THz detectors, for example as a power supply for distributed environmental sensor networks. Finally, we note that the MIG diode concept can be expanded to act as the emitter part of vertical hot electron transistors or graphene Base transistors, which are considered as three terminal devices operating up to the THz regime.[43–48] However, it is beyond the scope of this review to include them here in detail.

**3.3. Circuits**

The unique features of the MIG diodes have been utilized in many circuit applications such as power detectors,[30,49–51] mixers,[52,53] frequency multipliers,[54] and receivers.[55,56] In this section we explain the advantage of employing MIG diodes in these circuit applications.

*3.3.1. MIG in power detector circuits*

We have shown in Section 2 that the combination of low junction capacitance and low junction resistance of MIG diodes results in high cutoff frequencies.[30] In addition, the low junction resistance enables an exceptionally wide dynamic range when utilized in power detector circuits compared to other semiconductor materials.[57,58] Moreover, the high asymmetry of the DC characteristics of the MIG diode makes it an excellent candidate compared to state-of-the-art MIM diodes which have low junction resistance as well.[59] The high asymmetry results in a strong nonlinearity of the MIG diodes at zero-bias. This feature enables zero-bias power detector circuit topologies resulting in low noise operation which in turn improves signal sensitivity. These MIG diode properties have been exploited in a linear, single-diode power detector as shown in Figure 4a. The circuit was characterized on-wafer while providing external matching and external lowpass filtering.[30] The measured dynamic range of 40 dB (Figure 4b) outperforms not only CMOS-based detectors[58,60] but also III-V Schottky diode-based detectors.[57] The MIG diodes were further explored in more complex circuits with on-chip matching, in particular three matched MIG diodes in a *V*-band, zero-bias, linear-in-dB power detector scheme.[49,50] The detectors and circuits have been fabricated using CVD grown graphene on a glass substrate.[49] A micrograph of the chip with the circuit occupying 0.15 mm$^2$ is shown in Figure 4c. The circuits display linear-in-dB characteristics with a dynamic range of at least 50 dB. The measured responsivity reaches 15 V/W at 60 GHz and 168 V/W at 2.5 GHz with a sensitivity of < -50 dBm (Figure 4d).[49]

A wideband 0-70 GHz, zero-bias power detector using a three-stage distributed power detector is reported in[51]. The circuit integrates in the distributed structure the input matching together with the output lowpass filter. The number of stages has been determined with the help of a dedicated MATLAB model and optimizing the circuit based on a developed figure-of-merit to determine the optimum number of stages, taking into consideration the available graphene technology together with the MIG diode small-signal model. Figure 4e shows a micrograph of the chip with a distributed three-stage power detector circuit. The characterization results show a wideband RF input range from 0-70 GHz (Figure 4f). In addition to the measured signal sensitivity of 70 dBm, the dynamic range of the fabricated distributed detector is 75 dB. Earlier graphene-based power detectors rely on graphene field effect transistors (GFETs), but their inherent large gate capacitance limits the maximum achievable bandwidth to below 5 GHz.[49,61,62]

*3.3.2. MIG in mixer circuits*

The exploitation of the MIG diode in frequency conversion circuits has been reported both for a microwave single-MIG-diode configuration[52] and in a double-balanced upconverter architecture.[53] In [52], a single on-chip diode is employed together with an off-chip resistive power combiner, an off-chip input matching, and an output *I*-to-*V* 50-Ω buffer in a microwave downconversion mixer as shown in Figure 5a. The input RF and LO signals are fed to the power combiner and the output is fed to the input matching. The combined signal is fed to the on-chip MIG diode using on-wafer standard ground-signal-ground (GSG) probes with zero-bias voltage. The downconverted signal is then fed to the 50-Ω buffer. The measured conversion loss of this arrangement is 15 dB for an LO drive signal of +5 dBm. The resulting mixer performance (Figure 5b) outperforms all state-of-the-art GFET-based mixers employing both single [63,64] and double-balanced GFET configurations.[65] In addition, the measured performance of the single-MIG diode mixer outperforms also mixer circuits based on state-of-the-art MIM diodes.[66] The demonstrated performance of the single-MIG diode mixer combined with high yield provided by the developed large-scale CVD fabrication process, encouraged the assessment of MIG diodes in even more complex configurations, *i.e.*

using more devices in one circuit. A fully integrated wideband microwave upconversion mixer circuit has been reported in [53] employing four-MIG diodes and covering the entire X-band frequencies, *i.e.* from 6-12 GHz. The mixer circuit integrates four MIG diodes together with the input RF and LO baluns as well as the impedance matching circuits. The chip micrograph highlighting the mentioned parts of the circuit is depicted in Figure 5c. The circuit demonstrates a measured conversion loss as low as 10 dB and provides an RF-to-LO isolation of better than 25 dB (Figure 5d). The loss performance of this mixer sets a new record in terms of conversion loss compared to any reported mixer based on GFETs[63–65,67]. Furthermore, the circuit presents even comparable performance to commercial Schottky diode-based mixers.[68] As a conclusion, the cost efficient, mixer circuits based on CVD MIG diodes outperform all other graphene-based mixer circuits and demonstrate a tantamount performance compared to established semiconductor technologies.

*3.3.3. MIG as a frequency multiplier circuit*

Resistive frequency multiplier circuits based on MIG diodes have been investigated in [54]. In paricular, a microwave fully integrated balanced frequency-doubler circuit has been designed and fabricated employing a pair of MIG diodes. The circuit, as illustrated in Figure 6a, consists of an input RF balun for the frequency range of 3.5-7 GHz, a pair of matched MIG diodes, and an output combiner. The micrograph of the fabricated chip with the circuit is shown in Figure 6b. The circuit delivers an output frequency range from 7-13 GHz with an output power of -15.3 dBm and a conversion loss of 25.3 dB (Figure 6c and 6d). The performance of this circuit opens up the opportunity to realize higher-order graphene-based frequency multiplier circuits, with excellent performance theoretically expected regarding conversion losses. Such circuits are difficult to realize by using GFETs.[68,69]

*3.3.4. MIG in receiver circuits*

The realization of a conventional receiver consisting of a low-noise-amplifier (LNA) followed by a downconversion mixer using graphene technology has two challenges. The first is that the poor saturation of the GFETs results in low gain necessary to amplify the weak receive signal.[70,71] The second comes from the frequency limitation due to the lower postulated cutoff frequency ($f_T$) and maximum oscillation frequency ($f_{MAX}$) of present GFETs. Although possible, mixer-first receivers using GFETs provide limited performance due to the high conversion loss of GFET-based mixers originating from their poor switching performance. However, receivers can also be built using diodes in general and MIG diodes in particular. MIG diodes are advantageously employed in two receiver architectures other than the conventional topologies, such as heterodyne, homodyne or mixer first. The first approach is the sixport receiver topology which represents a solution for the second challenge mentioned above. A sixport receiver as shown in Figure 7a consists of passive couplers and combiners that add the received RF signal and the reference LO signals with different phases and feed these additions to four matched, square-law power detectors, where MIG diode-based power detectors are used because of their good RF power detection capabilities.[55] The sixport receiver is designed for an input RF frequency range from 2.1-2.7 GHz utilizing lumped-element implementation for the passive couplers and power combiner. The micrograph of a chip with a complete receiver is shown in Figure 7b occupying only an area of 4.8 mm². The receiver capabilities are verified by the reception of a 20-Mbps QPSK signal at 2.45 GHz with −15 dBm of modulated RF input power strength while the LO power was set to 0 dBm (Figure 7c). The measured conversion gain for a single tone is -7 dB. Although this sixport receiver approach has proven to be able to solve the second challenge, *i.e.* low $f_T$ and $f_{MAX}$, of graphene technologies, the positive conversion gain at high frequencies challenge is not addressed with this approach. Conversely to transconductance amplification widely used today, parametric or transreactance amplification has been used widely since the 1950s[72]. The basic design equations describing this approach have been reported in.[73,74] The main advantage of parametric amplification is that it can be used to provide at the same time frequency conversion and amplification. Utilizing the parametric approach for frequency down conversion using conventional Schottky varactors and later CMOS devices is challenging due to the capacitance/voltage (C/V) relation of these varactors as shown in Figure 8a. A practical implementation limitation comes from the required high power LO signal at frequencies higher than the received RF frequency. This is known as a lower-sideband parametric down conversion. The exploitation of the unique C/V characteristics of the MIG diodes illustrated in Figure 8b was introduced for the first time in [56]. The main advantage of employing MIG diodes according to findings in [56] is that due to their C/V characteristics they intrinsically perform even harmonics generation and cancellation of the fundamental which solves the limitation of the parametric down conversion discussed above.

## 4. Optoelectronic Devices

The conversion of light to electricity is of fundamental importance in many technological applications such as imaging, optical communications, night-vision, motion detection and solar cells. Several key figures-of-merit can be used to quantify the performance of a photo detector, such as responsivity ($R_{ph}$, measured in A/W), noise equivalent power, noise equivalent irradiance, and the time response. These parameters respectively define 1) the ratio between the photocurrent (the difference of the measured current when the light illuminates the detector and the current in the dark condition) to the incident light power, 2) the lowest detectable light power, 3) the lowest detectable light intensity, and 4) the time a photodetector takes to respond to an optical input.[21] Further relevant figures of merits are the external quantum efficiency (the ratio between free charge carriers and incident photons), the internal quantum efficiency (the ratio between free charge carriers and absorbed photons), and the detectivity (the inverse of the noise equivalent power, normalized to the detector's area and bandwidth).[75,76]

Currently, the optoelectronics and photonics market is dominated by complementary metal-oxide-semiconductor (CMOS) technology developed on Si in the visible spectrum owing to advanced and cost effective processing technology.[77] However, due to the band gap of Si of ~ 1.1eV, infrared absorption of Si is limited and III−V semiconductors like Germanium (Ge) or Indium Gallium Arsenide (InGaAs) are replacing the role of Si for infrared detection, especially at the communication wavelengths of approximately 1.5 μm.[3,77,78] However, complex and costly growth of these III−V semiconductor materials and challenging CMOS fabrication line integration for mass production still leave room for further advantageous alternatives.[79] Graphene,[12] owing to its unique properties such as high mobility at room temperature[80,81] and broadband optical absorption[31,82] as well as its back-end-of-line (BEOL)-compatible CMOS integration scheme through transfer,[83,84] appears to be an ideal candidate for photo detection and 3D CMOS integration.[85–87] However, weak absolute optical absorption of graphene limits its photoresponsivity in many applications. Nevertheless, one can combine the advantages of graphene with those of other semiconductor materials like Si or with photoactive materials like quantum dots (QD) to enhance the photoresponse.

Graphene field effect transistor (GFET) based photodetectors with $R_{ph}$ up to $10^6$ A/W in the visible spectrum and $10^8$ A/W in the near infrared have been reported as strong candidates for replacing III-V semiconductor based infrared photo detectors.[88,89] However, the detection mechanism of these photodetectors relies highly on absorbing layers such as QDs,[88] dyes like J-Aggregates[90] or other two-dimensional materials like molybdenum disulfide ($MoS_2$)[89,91] coated/transferred on graphene. Upon illumination, light is absorbed by the active absorbing layer, which in turn induces a shift in the chemical potential of graphene and thus changes the resistance of the graphene channel. This can be directly measured as a photocurrent.[88–91] However, GFET-based photodetectors suffer from high dark currents due to the semi-metallic nature of graphene. The devices are always "on", which leads to high power consumption during operation.[11] Semiconducting photodetectors made from Ge and InGaAs provide much lower dark current levels of a few nanoamperes, while showing responsivities up to 1 A/W.[92–94] MoS$_2$-based detectors present another alternative for infrared photodetection, either functionalized through absorption layers similar to graphene[95] or directly through defect band absorption.[96,97] They yield high $R_{ph}$ up to ∼ 6 × 10$^5$ A/W, while keeping the dark current at below 1 μA. However, these devices so far show a limited time response $\tau_{res}$ of approximately 0.3 s, which prohibits applications in real-time IR imaging.[21] Here, we focus on photodiodes with graphene as one of the functinal components.

**4.1. Graphene-Si Schottky Diodes**

Graphene/Si Schottky heterojunctions photodiodes have been studied intensively as their fabrication is quite compatible with conventional semiconductor technology. [36,98–106] In these photodiodes, the extraction of photogenerated charge carries is governed by a graphene/Si Schottky contact (Figure 9a). Studies on devices with CVD-grown graphene on lightly doped n-type Si have shown that when the device is reversely biased, photons mainly generate electron-hole pairs in the Si, with holes being collected by the graphene (Figure 9b). The photogeneration and the transport of charge carriers happens mainly in the depletion layer of the Si and only those carriers that are able to reach the graphene before recombining contribute to the photocurrent. Therefore, the responsivity is limited to the maximum responsivity of the n-Si substrate and to the carrier lifetime, for energies above its bandgap. This has been confirmed by spectrally resolved photocurrent measurements that show the well-known fingerprint of silicon photodetection in the visible spectrum (Figure 9c). Nevertheless, some responsivity is also recorded in the IR region (below the bandgap of Si), approximately three orders of magnitude lower than the maximum responsivity in the visible range, which is attributed to the absorption in the graphene (inset of Figure 9c).[36]

Improving the responsivity of graphene/Si photodiodes beyond the state of the art in conventional silicon has been targeted and achieved by several groups. Graphene/Si diodes have shown high responsivity up to $10^7$ A/W when biased in high-gain mode, which is much higher than the maximum responsivity of Si.[107] The result is explained by the underlying photo response mechanism, termed quantum carrier reinvestment, in which photogenerated carriers of n-type Si are "borrowed" to graphene and reinvested several times in the external circuit during their lifetime, which can exceed a millisecond, resulting in ultrahigh quantum gain values. Impact ionization has also been suggested as the origin of high responsivity and high quantum efficiency.[108] Here, the graphene is placed across $SiO_2$ over a Si trench to form a device consisting of a graphene−Si heterojunction in parallel with a graphene/insulator/Si (GIS) field effect structure, and a graphene / air gap / silicon part at the transition of the two regions. The GIS region leads to an inversion channel under certain conditions, termed 2D electron gas in the work, where carrier multiplication is thought to occur and lead to high photocurrents and quantum efficiencies well above unity. Similar structures of graphene / Si Schottky diodes in parallel with GIS structures have been investigated with the scanning photocurrent microscopy technique, which provides the spatial distribution of the photocurrent generation in devices and capacitance voltage measurements.[109,110] Similar to [108], considerable photocurrents have been observed above a certain reverse bias threshold underneath the isolating GIS region of graphene on $SiO_2$ adjacent to the Schottky junction (Figure 9d and 9e). The formation of an inversion layer, similar to the channel in a metal oxide semiconductor field effect transistor, has been confirmed through electrostatic simulations. This inversion layer provides a low resistance path for the generated charge carries and passivates surface states in $SiO_2$, enhancing the efficient collection of photocurrent. This mechanism has been exploited in diodes with specifically tailored interdigitated Schottky and GIS regions, where the GIS areas are used for efficient carrier generation and the Schottky areas are used for both carrier generation and carrier collection in the graphene (Figure 10a).[111] Figure 10b shows a photograph of a wire-bonded chip with several such photodiodes. The interdigitated devices show higher responsivity of up to 0.6 A/W than commercial silicon p-n diodes (Figure 10c) and an external quantum efficiency approaching 100% (Figure 10d). A similar interdigitated concept has been used to demonstrate enhanced IR responsivity, which can be attributed to locally enhanced electric fields at the edges between the Schottky and GIS regions.[112] Finally, arrays of Si nanotips have been proposed as support substrates for graphene/Si Schottky diodes.[113] Here, the nanotip surface enhances light collection due to multiple reflections and the tip-enhanced field further supports the separation of photogenerated carriers with internal gain due to impact ionization. These structures result in responsivity up to 3 A/W, which is one to two orders of magnitude higher than that reported for planar graphene/Si junctions. [36,99,109] There are many reports of Schottky and p-n diodes with semiconducting 2D materials like $MoS_2$, platinum diselenide ($PtSe_2$) and others, as well as purely 2D heterostructure photodetectors, including the IT sensitive black phosphorous. [114–118,97] However, their detailed review is beyond the scope of this paper and may be found elsewhere.[21,119,120]

**4.2. QD/MIG Photodiodes**

The combination of the MIG diode concept with IR sensitive quantum dots has been proposed to realize IR photodetectors that show orders of magnitude lower dark current levels compared to GFET-based approaches while keeping the responsivity level a few orders of magnitude higher than in Ge- and InGaAs-based detectors.[121] This has been achieved by photo sensitization of a MIG diode structure with colloidal quantum dots (CQDs) as the photo-absorbing layer. A schematic of a photodetector is shown in Figure 11a. After the diode fabrication on quartz substrate with a Titanium metal contact, a 5-nm $TiO_x$ insulator and graphene, lead sulfide (PbS) CQDs with a first absorption peak at approximately 1625 nm were spin-casted on graphene for photo sensitization.[121] The working principle of the MIG/QD photo detector depends on the manipulation of the chemical potential of the graphene caused by light absorption in the QD layer, similar to other reported graphene-based hybrid photodetectors. Light illumination on the devices generates electron−hole pairs in the CQD layer and a fraction of these carriers are then transferred to the graphene (Figure 11b). As a result, the chemical potential of graphene increases and results in a photocurrent through the MIG diode.

Figure 12a shows the device responsivity as a function of $V_{bias}$ at different light power densities, $PD$, at infrared light of 1625 nm. A detectable photocurrent was present at $V_{bias} \lesssim -0.74$ V and $V_{bias} \gtrsim 0.34$ V. In between these values, the photocurrent was below the noise floor of measurement setup. The maximum measured $R_{ph}$ reaches values as high as $\sim 3 \cdot 10^4$ A/W at 1 V $V_{bias}$. However, this bias voltage causes high current levels in the diode even without illumination due to the exponential dependency of the current on the diode barrier height, i.e. high dark currents in the range of tens of microampere.[30] Figure 12b shows the measured responsivity versus the dark current densities. The optimal operating $V_{bias}$ was determined as $\sim 0.5-0.8$ V , where the dark-current is in the range of hundreds of nanoamperes. Figure 12b also compares the measured data to the values of reported Ge-, InGaAs-, and GFET-based photodetectors. The "sweet spot" for the MIG/QD diodes at optimal $V_{bias}$ shows two to three orders of magnitude lower dark currents than GFETs, and at the same time two orders of magnitude improvement in responsivity compared to Ge and InGaAs photodetectors.[92,94,122] The time response, $\tau_{res}$, of the MIG/QD-based photodetectors was reported as 0.14 ms, which is similar to that of GFET-based detectors, and about three orders of magnitude faster than high-responsivity photodetectors relying on MoS$_2$[91,95,96]. Another figure of merit to determine the performance of photo detectors is noise equivalent irradiance (NEI), the lowest detectable input power density. Figure 12c shows the measured power density at different $V_{bias}$. Regardless of the bias voltage, the photo response increases linearly between 0.03 and 1W/m$^2$. Thus, the NEI of the device was determined as 0.03W/m$^2$.

Based on these values, the extracted $R_{ph}$ at the optimal power density close to the NEI was 70 A/W in the infrared at $V_{bias}$ = 0.5 V, while the dark current was 700 nA ($J_{dark} \sim 0.3$ A/cm$^2$). These numbers confirm the potential of this device concept for efficient photodetection in the infrared wavelength range. Moreover, the combination with MIG diode-based electronic readout circuits could lead to low cost, low power graphene-based infrared image sensors as an alternative for state-of-the-art Ge and InGaAs technologies.

## 5. Conclusions

2D/3D heterostructures are comparably simple devices that allow integration of 2D materials into conventional semiconductor and thin-film technology. Experimental data demonstrates performance advantages of both metal-insulator-graphene and metal-semiconductor Schottky diodes over conventional technologies in specific electronic and optoelectronic applications. Circuit applications incorporating graphene diodes as high-frequency elements demonstrate their capability to advance technology for the Internet of Things, analog THz signal processing, and energy harvesting. Graphene diode-based photodetectors show improved figures-of-merit for responsivity, infrared sensitivity and/or low power operation, with potential for graphene-based image sensors suitable for real time imaging, which may be an alternative to Ge-and InGaAs-based technologies. Among the main challenges towards the commercialization of graphene-based diodes are the lack of growth and transfer technologies for graphene and the integration technology into conventional semiconductor platforms. Progress in the quality and integrability of large-area grown graphene is expected to improve over the coming years[123] and enable further improvements in key device characteristics such as cutoff frequency, current levels or responsivity. This may pave the way for graphene to become an integral part of future commercial electronic and optoelectronic components.

**Acknowledgements**

The authors acknowledge German Research Foundation (DFG) under the project GLECS II (No. NE1633/3-2), HiPeDi (No. WA 4139/1-1), the German Ministry of Education and research (BMBF) under the project GIMMIK (03XP0210), and the European Commission under the Horizon 2020 projects Graphene Flagship (No. 785219 and No. 881603), ULISSES (825272), QUEFORMAL (829035), WiPLASH (No. 863337) and 2D-EPL (952792).

Received: ((will be filled in by the editorial staff))

Revised: ((will be filled in by the editorial staff))

Published online: ((will be filled in by the editorial staff))

**Table 1.** Comparison of the static performance of state-of-the-art MIM and MIG diodes. Parameters such as current level *I* at bias of 1 V, current density *J*, asymmetry $f_{ASYM}$, nonlinearity $f_{NL}$, responsivity $f_{RES}$ are used for benchmarking. MIG diodes in general show higher current level and higher $f_{ASYM}$, $f_{NL}$, $f_{RES}$, given similar barrier thickness.

| | I (at 1 V) | J (A/cm$^2$) | $f_{ASYM}$ | $f_{NL}$ | $f_{RES}$ (V$^{-1}$) |
|---|---|---|---|---|---|
| Nb/Nb2O5(5 nm)/Pt[37] | 128 uA | 2 | 9.8 | 8.2 | 16.9 |
| Nb/Nb2O5(15 nm)/Pt[7] | 175 nA (at 0.5 V) | | 1500 | 4 | 20 |
| Ti/TiO2/bilayer graphene[27] | 10 nA | 0.1 | 9000 | 8 | 10 |
| Ti/TiO2/graphene (max)[30] | 10.5 uA | 7.5 | 520 | 15 | 26 |
| 1D-MIG (max)[42] | 1.6 mA | 7.5x10$^6$ | 65.7 | 4.5 | 18.5 |

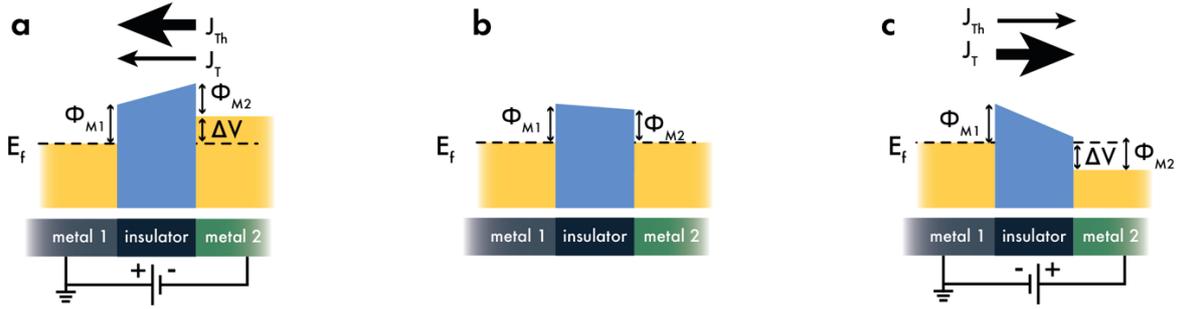
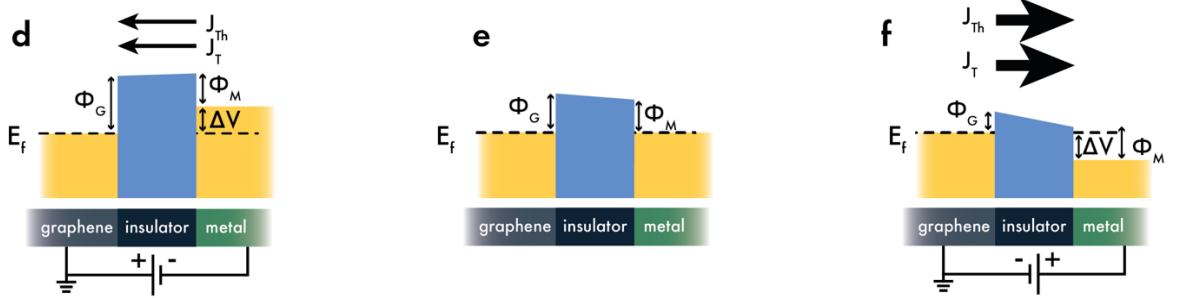

**Figure 1.** (a) – (c) Working principle of metal-insulator-metal diodes for (a) reverse bias, (b) zero bias and (c) forward bias. $\Phi_{M1}$ and $\Phi_{M2}$ represent the difference between the electron affinity of the insulator and the work function of metal 1 and 2, respectively. (d) – (f) Working principle of metal-insulator-graphene diodes for (d) reverse bias, (e) zero bias and (f) forward bias. $\Phi_G$ is bias-dependent. $E_f$: Fermi level, $J_T$: tunneling current, $J_{Th}$: thermionic current, $\Delta V$: applied bias, $\Phi_G$: graphene work function, $\Phi_M$: metal work function, $\Phi_{M1}$: metal 1 work function, $\Phi_{M2}$: metal 2 work function.

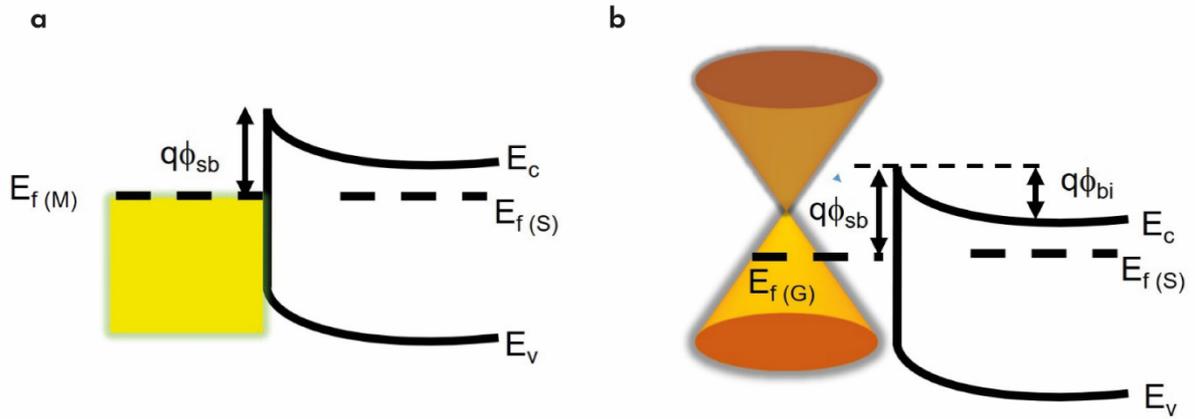

**Figure 2**. (a) Schematic band diagram of a metal-semiconductor junction, and (b) a graphene-semiconductor (n-type) junction under zero bias condition. f$_{sb}$ is variable in case of the graphene-semiconductor junction due to the bias dependent Fermi energy level of graphene. E$_c$: conduction band edge, E$_f$(G): graphene Fermi level, E$_f$(M): metal Fermi level, E$_f$(S): semiconductor Fermi level, E$_v$: valence band edge, q: elementary charge, $\Phi_{bi}$: built-in potential, $\Phi_{sb}$: Schottky barrier height.

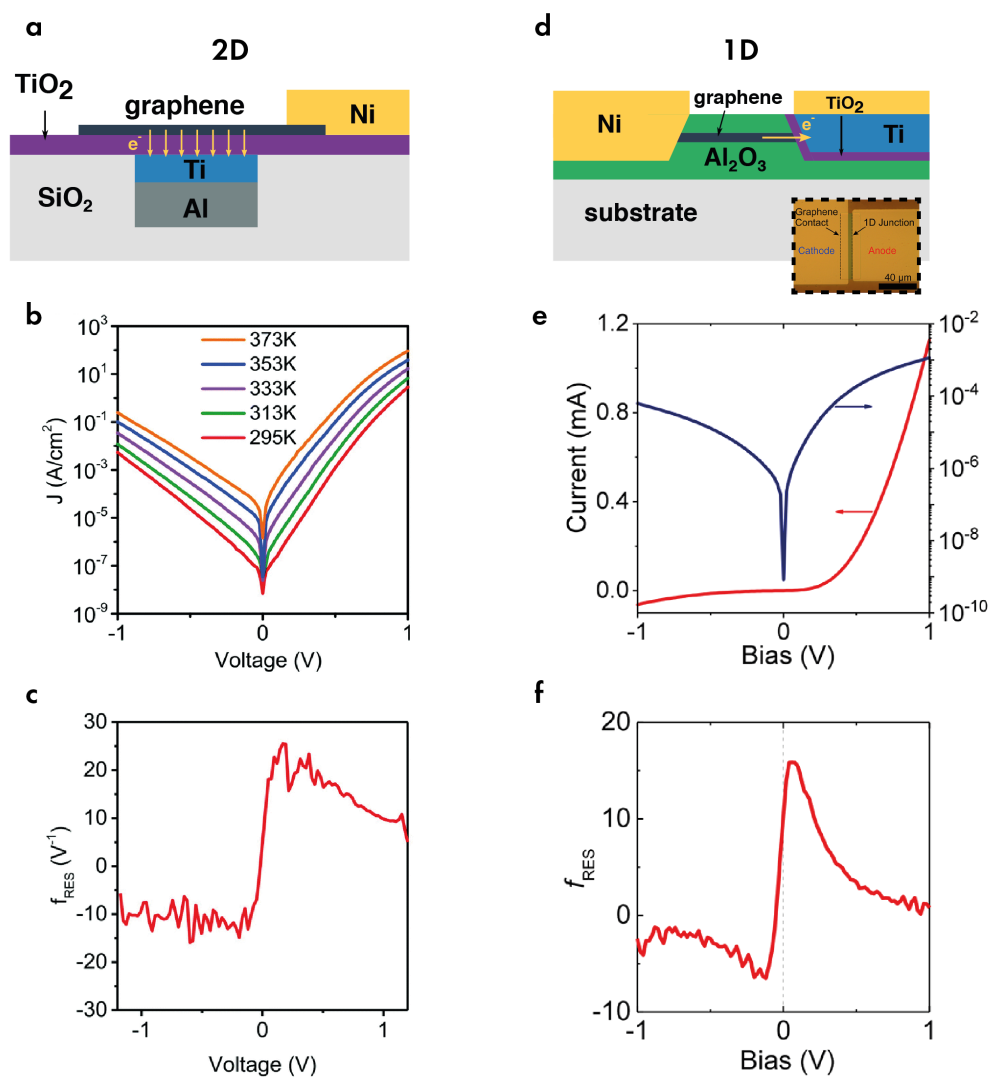

**Figure 3.** (a) Schematic cross section of a 2D MIG diode. (b) Current density-voltage curve at different temperatures. (c) Responsivity of the 2D MIG diode as a function of applied bias voltage. (d) Schematic cross section of a 1D MIG diode. The inset shows an optical micrograph of a 1D MIG diode. (e) Linear and logarithmic current-voltage characteristics of the 1D MIG diode. (f) Responsivity of the 1D MIG diode as a function of applied bias voltage.[14] $f_{Res}$: responsivity, J: current density. (b), (c) Reproduced with permission.[30] Copyright 2017, The Royal Society of Chemistry. (e), (f) Reproduced with permission.[42] Copyright 2019, American Chemical Society.

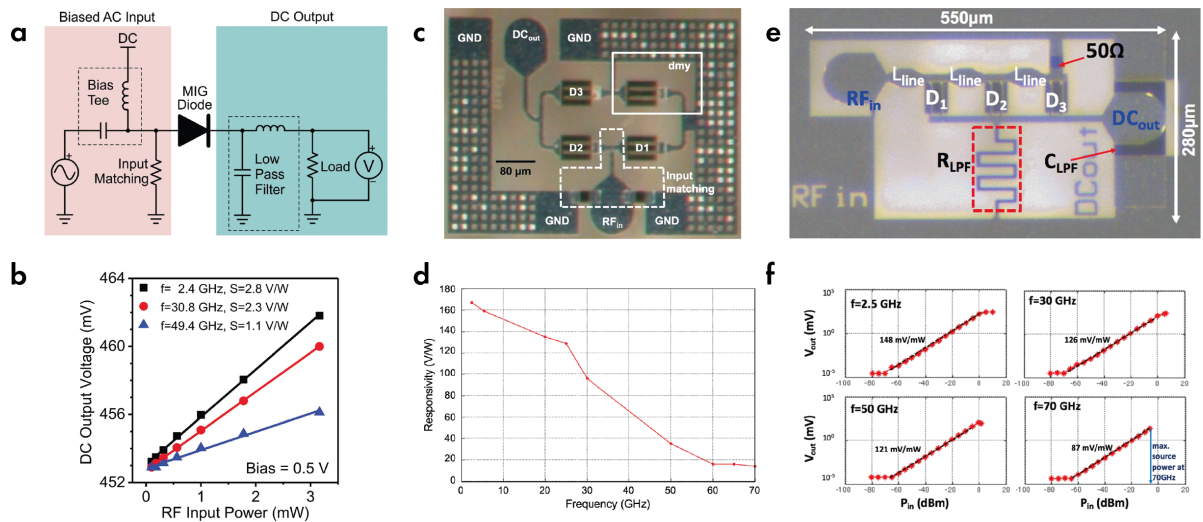

**Figure 4**. MIG diode-based power detector circuits: (a) Measurement setup for the single MIG diode power detector, and (b) measured DC output voltage as a function of RF input power. (c) Chip micrograph of the linear-in-dB power detector, and (d) the measured responsivity as a function of frequency. (e) Chip micrograph of the distributed power detector, and (f) the measured detected voltage as a function of input power at different frequencies. $C_{LPF}$: low pass filter capacitance, D1: diode 1, D2: diode 2, D3: diode 3, $DC_{out}$: DC output, GND: ground, IF: intermediate frequency, $IF_{out}$: intermediate frequency output, $L_{line}$: line inductance, $LO_{in}$: Local oscillator input, $RF_{in}$: RF signal input, $R_{LPF}$: low pass filter resistance. (a), (b) Reproduced with permission.[30] Copyright 2017, The Royal Society of Chemistry. (c) - (f) Reproduced with permission.[49,51] Copyright 2018, IEEE.

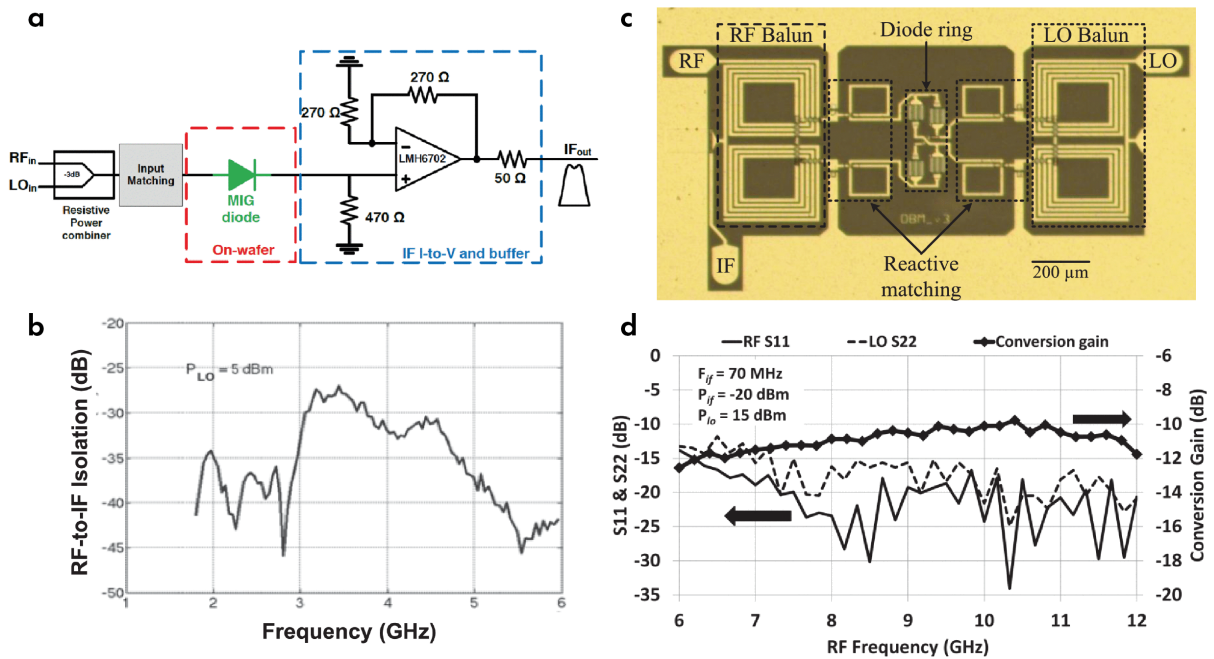

**Figure 5**. MIG diode-based mixer circuits: (a) Measurement setup for the single MIG diode microwave downconversion mixer, and (b) the measured RF-to-IF isolation. (c) Chip micrograph of the double-balanced upconversion mixer, and (d) the measured conversion loss and S11 and S22 parameters. IF: intermediate frequency, LO: local oscillator, $P_{LO}$: local oscillator power. (a), (c) Reproduced with permission.[124] Copyright 2019, IEEE. (b), (d) Reproduced with permission.[52,53] Copyright 2018, IEEE.

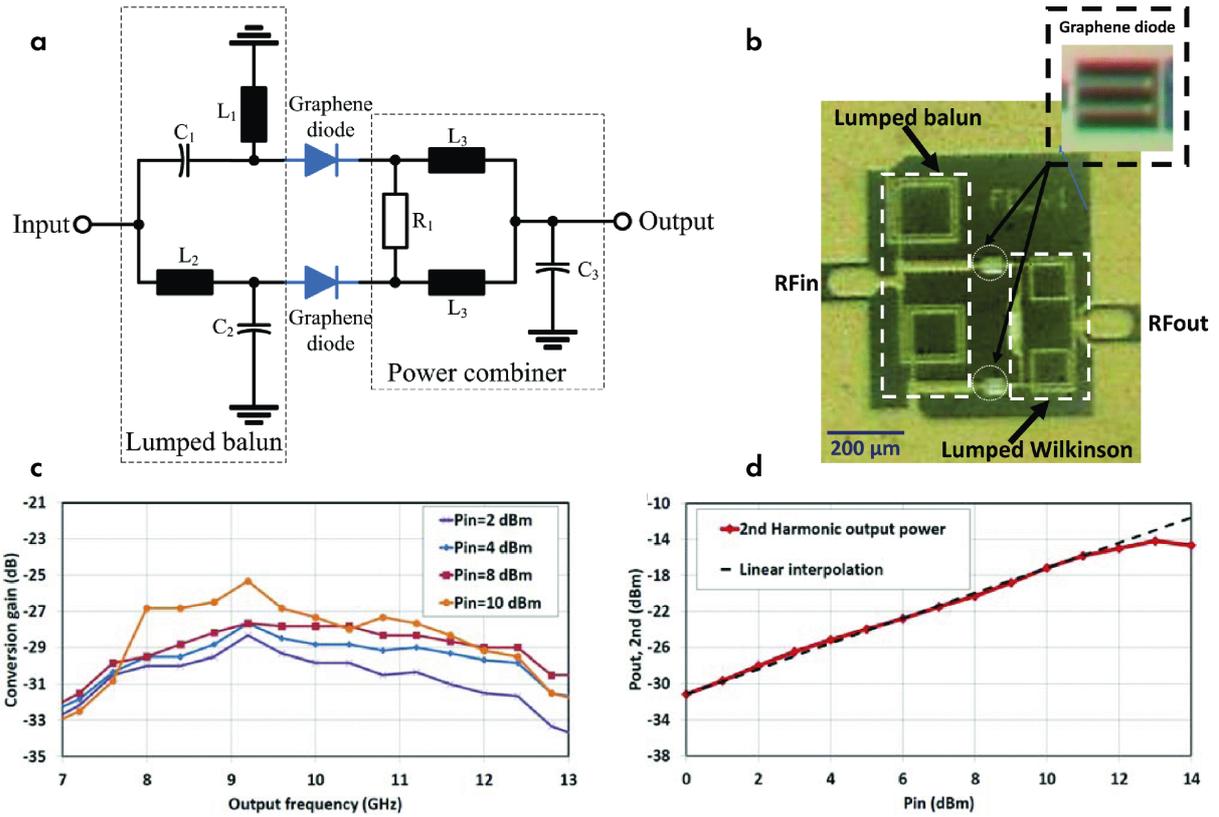

**Figure 6**. MIG diode-based frequency doubler: (a) circuit schematic, and (b) chip micrograph. (c) shows the conversion loss over frequency and (d) the output power as a function of input power. C1: capacitance 1, C2: capacitance 2, C3: capacitance 3, L1: inductance 1, L2: inductance 2, L3: inductance 3, R1: resistance 1, $RF_{in}$: RF signal input, $RF_{out}$: RF signal output. Reproduced with permission.[54] Copyright 2019, IEEE.

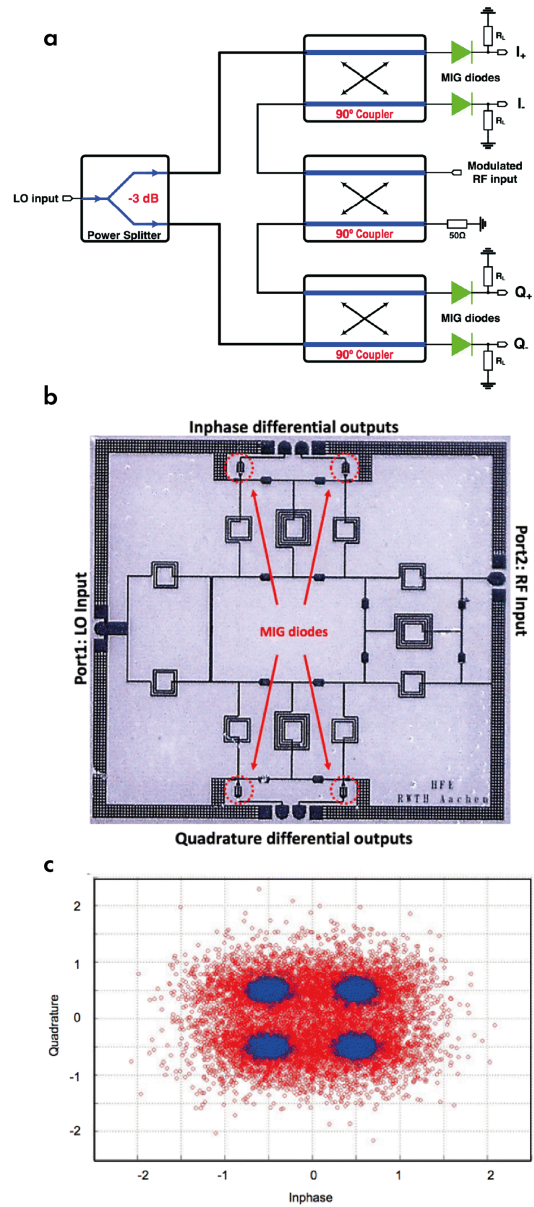

**Figure 7**. Sixport receiver: (a) block diagram of the sixport receiver, (b) Chip micrograph of the fabricated sixport receiver, and (c) constellation diagram of a QPSK signal before calibration (red) and after calibration (blue). LO: local oscillator. Reproduced with permission.[55] Copyright 2017, The Royal Society of Chemistry.

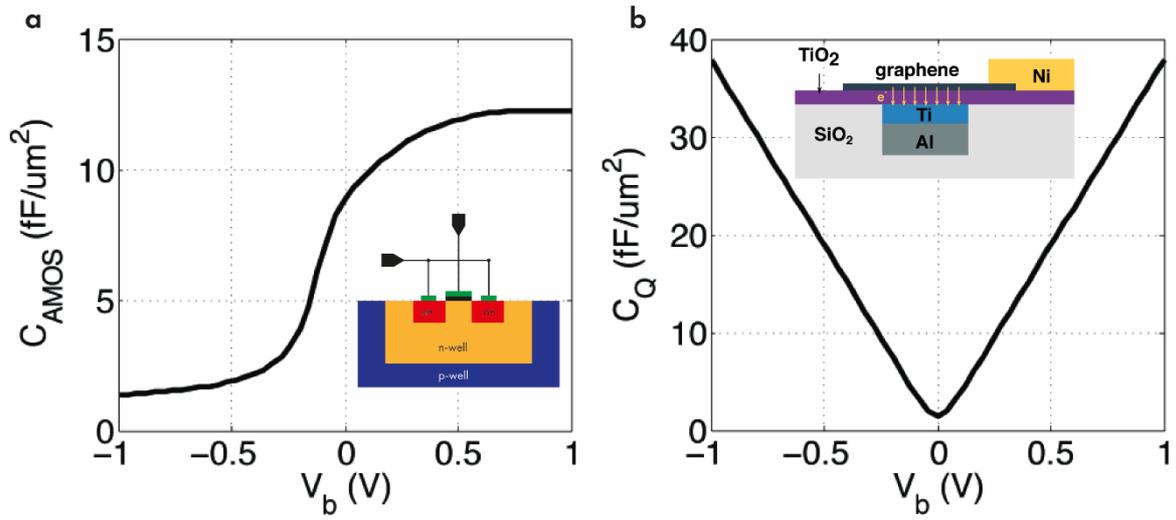

**Figure 8**. Capacitance-voltage characteristics with cross sections shown in the inset: (a) Conventional A-MOS varactor, and (b) graphene quantum capacitance. $C_{AMOS}$: accumulation-mode MOS capacitance, $C_Q$: quantum capacitance, $V_b$: bias voltage. Reproduced with permission.[56] Copyright 2019, IEEE.

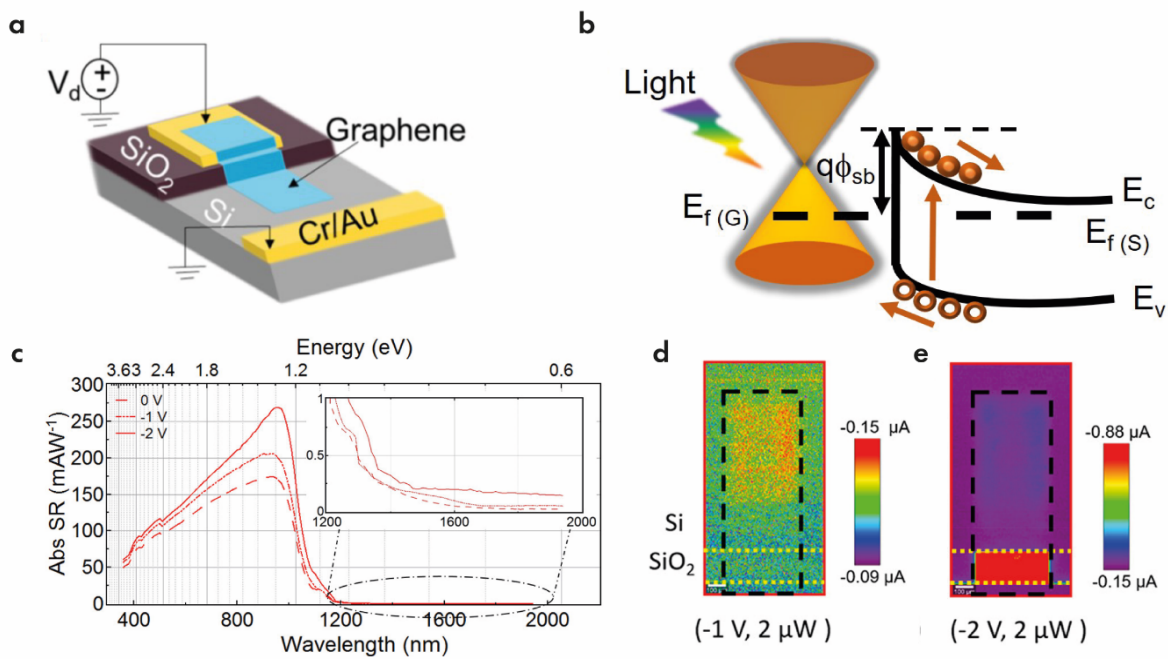

**Figure 9**. (a) Schematic of a graphene-Si junction in photodiode mode. (b) Band diagram showing the graphene-Si junction under reverse bias in illuminated conditions. Charge carriers are generated in Si with holes being collected by graphene. (c) Bias dependent spectral responsivity (SR) of the device shown in (a) with most of the response coming from Si above its bandgap of 1.1 eV. Inset shows a small yet measurable response below the band gap of Si which is attributed to graphene. (d) and (e) show the scanning photocurrent measurements performed on the device shown in (a) which prove that most of the photocurrent is generated below graphene/SiO2 region in the device. Dotted rectangle indicates area covered by graphene. Abs SR: absolute spectral responsivity, $E_c$: conduction band edge, $E_f(G)$: graphene Fermi level, $E_f(S)$: semiconductor Fermi level, $E_v$: valence band edge, q: elementary charge, $V_d$: drain voltage, $\Phi_{sb}$: Schottky barrier height. (a) (c) Reproduced with permission.[36] Copyright 2016, Elsevier. (d) Reproduced with permission.[109] Copyright 2017, American Chemical Society.

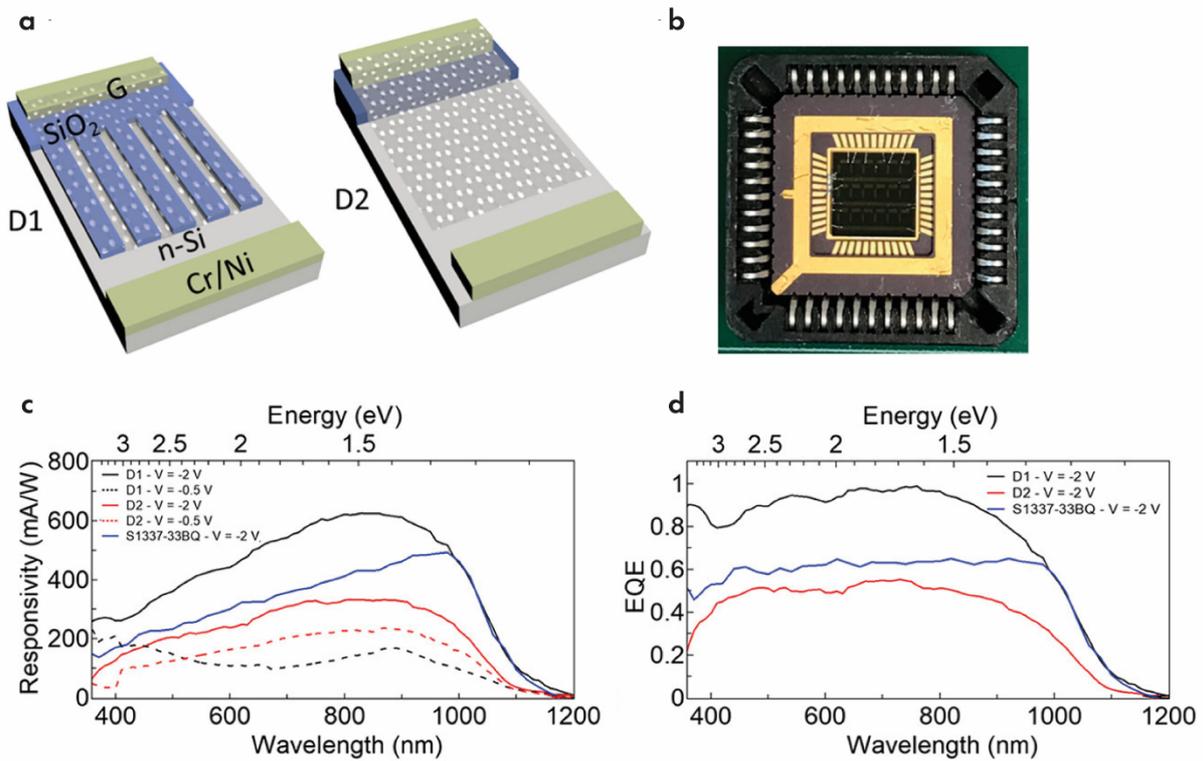

**Figure 10**. (a) Schematic of a G/Si photodiode with (D1) and without (D2) interdigitated G/SiO2 and G/Si regions. (b) Photograph of a packaged photodiode chip. (c) Responsivity curves of D1, D2 and a commercial Si photodiode under different reverse bias conditions. Device D1 exhibits an improved responsivity comparative to other devices. (d) EQE of devices D1, D2 and a commercial Si photodiode at a reverse bias of -2 V. D1 is found to outperform even the commercial Si photodiode with an maximum EQE of around 98%. D1: diode 1, D2: diode 2, EQE: external quantum efficiency, G: ground. Reproduced with permission.[111] Copyright 2019, American Chemical Society.

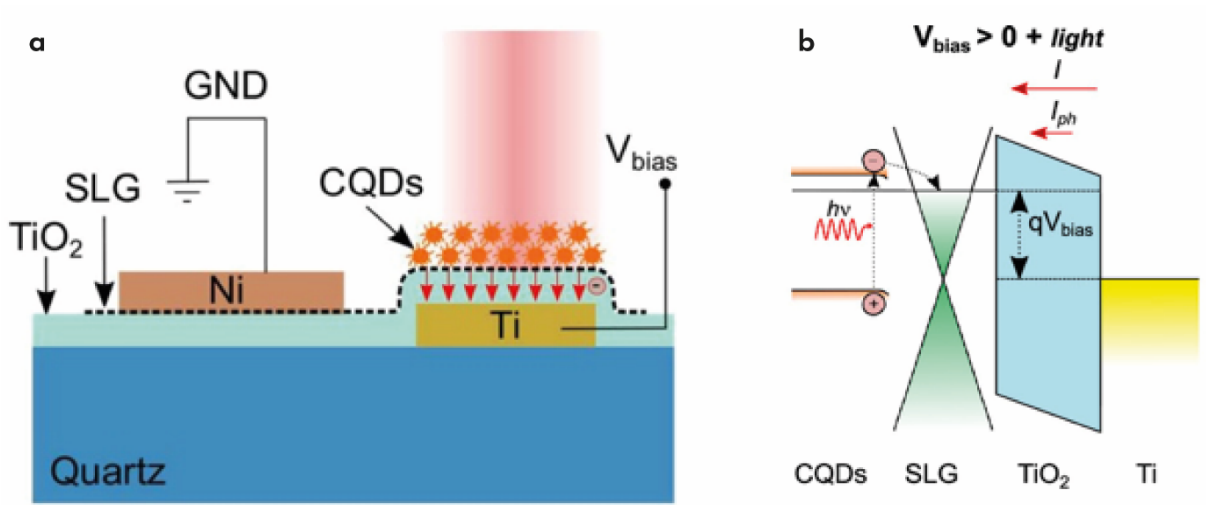

**Figure 11**. (a) Schematic of the metal-insulator-graphene /CQD photodetector. (b) Sketch of the band diagram under light illumination. CQDs: colloidal quantum dots, I: current, $I_{ph}$: photocurrent, q: elementary charge, SLG: single-layer graphene, $V_{bias}$: bias voltage. Reproduced with permission.[121] Copyright 2020, American Chemical Society.

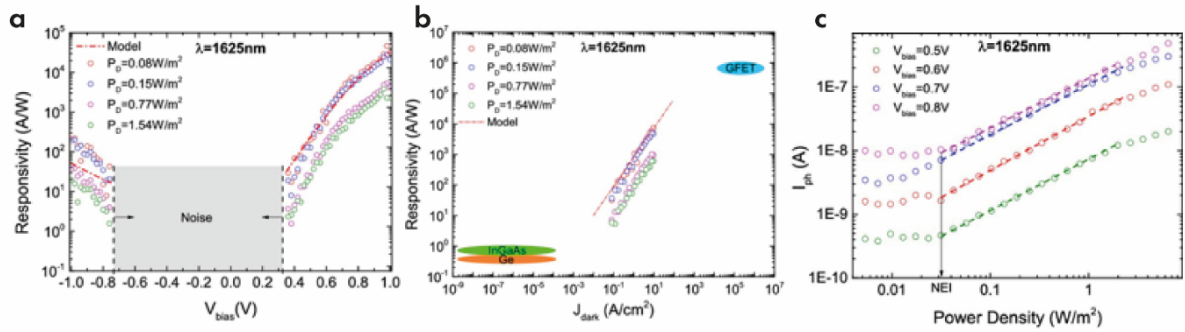

**Figure 12**. (a) Device responsivity upon light illumination at λ=1625nm. (b) Responsivity $R_{ph}$ as a function of dark current density $J_{dark}$ flowing in the device without light illumination. (c) Photocurrent measurement against varying light power densities at different bias voltages. $I_{ph}$: photocurrent, $J_{dark}$: dark current, $P_D$: power density, $V_{bias}$: bias voltage, $\lambda$: wavelength. Reproduced with permission.[121] Copyright 2020, American Chemical Society.